\documentclass[amsmath,amssymb,prl,aps,mathbbm,superscriptaddress,twocolumn]{revtex4-1}
\usepackage{amsmath,amsfonts,amssymb,amsthm,graphics,graphicx,epsfig,times,bbm}
\usepackage[colorlinks=true,citecolor=blue,linkcolor=magenta]{hyperref}
\usepackage[usenames]{color}
\usepackage{graphicx}
\usepackage{subfigure}
\usepackage{amsmath}
\usepackage{epsfig,times}
\usepackage{dcolumn}
\usepackage{bm}
\usepackage{color}
\usepackage{verbatim}
\usepackage{epstopdf}
\usepackage{amssymb}
\usepackage{amstext}
\usepackage{latexsym}
\usepackage{hyperref}
\usepackage{amsfonts}
\usepackage{psfrag}
\usepackage{verbatim}
\usepackage{xcolor}

\usepackage{mathptmx} 

\usepackage[colorlinks=true,citecolor=blue,linkcolor=magenta]{hyperref}

\def\s{{\rm\bf \sigma}}

\def\p{{\rm\bf p}}
\def\q{{\rm\bf q}}

\def\U{{\rm U}}

\def\la{\langle}
\def\ra{\rangle}
\def\om{\omega}
\def\Om{\Omega}

\newcommand{\beqa}{\begin{eqnarray}}
\newcommand{\eeqa}{\end{eqnarray}}

\begin{document}

\title{Shortcuts to adiabaticity by counter-diabatic driving }
\author{Adolfo del Campo}
\affiliation{Theoretical Division,  Los Alamos National Laboratory, Los Alamos, NM 87545, USA}
\affiliation{Center for Nonlinear Studies,  Los Alamos National Laboratory, Los Alamos, NM 87545, USA}
\begin{abstract}

The evolution of a system induced by counter-diabatic driving mimics the adiabatic dynamics without the requirement of slow driving. 
Engineering it involves diagonalizing the instantaneous Hamiltonian of the system and results in the need of auxiliary non-local interactions for matter-waves. 
Here  experimentally realizable  driving protocols  are found for a large class of single-particle, many-body,  and non-linear systems without demanding the spectral properties as an input. The method is applied to the fast decompression of Bose-Einstein condensates in different trapping potentials.

\end{abstract}


\maketitle


The development of new methods to induce  adiabatic dynamics is key to the progress of quantum technologies \cite{CZ12}.
The emergent field of shortcuts to adiabaticity  aims at designing non-adiabatic protocols which reproduce the same target state 
that would result in a strictly adiabatic dynamics. In the last few years, a great deal of attention has been devoted to this goal,  
and theoretical efforts \cite{DR03,Berry09,MN10,Chen10,Muga10,Stefanatos10,Hoffmann11,delcampo11b,RC11,Torrontegui11,DB12,Stefanatos13} have been accompanied by a remarkable experimental progress \cite{Schaff1,Schaff2,ions1,ions2,expCD1,expCD2}. 
Among the variety of techniques developed, counter-diabatic driving (CD) \cite{DR03}, also known as transitionless quantum driving \cite{Berry09}, stands out as a technique 
to control and engineer the fast evolution of a system mimicking adiabaticity. Indeed, it is tantamount to induce a ``fast motion video of the adiabatic dynamics''.
 Consider a time-dependent Hamiltonian $\hat{H}(t)$ with instantaneous eigenvalues $\{\varepsilon_n(t)\}$ and eigenstates $\{|n(t)\ra\}$.
Whenever $\hat{H}(t)$ is slowly-varying, the dynamics for the $n$-th eigenstate in the adiabatic approximation reads
\beqa
\label{adiabsol}
|\psi_n(t)\ra=e^{-i\int^tdt'\varepsilon_n(t')}e^{-\int^tdt'\la n|\partial_{t'}n\ra }|n(t)\ra,
\eeqa
which includes both the dynamical phase and the geometric phase.
The central goal  of CD is to find a Hamiltonian $\hat{H}_{\rm CD}$ for which the adiabatic approximation to $\hat{H}$ 
becomes the exact solution of the time-dependent Schr\"odinger equation for $\hat{H}_{\rm CD}$.
Direct construction of the time evolution operator 
\beqa
\hat{U}_{\rm CD}(t,t'=0)=\sum_n |\psi_n(t)\ra\la n(0)|,
\eeqa
allows one to derive the form of $\hat{H}_{\rm CD}$, 
\beqa
\label{hCD}
\hat{H}_{\rm CD}&=&i\hbar\hat{U}_{\rm CD}(t)^{\dag}\partial_t\hat{U}_{\rm CD}(t)=\hat{H}+\hat{H}_1,\\
\hat{H}_1&=& i\hbar\sum_n(|\partial_tn\ra\la n|-\la n|\partial_t n\ra |n\ra\la n|).
\eeqa
It follows that $\hat{H}_1$ is the auxiliary term required to implement this type of shortcut. 
Note that other choices of the phase in Eq. (\ref{adiabsol}) are possible. An example of relevance is the dynamics $|\phi_n(t)\ra=
e^{-i\int^tdt'\varepsilon_n(t')}|n(t)\ra$, associated with the unitary time-evolution operator  $\hat{U}_{\rm CD}'(t,t'=0)=\sum_n |\phi_n(t)\ra\la n(0)|$, and generated by the counter-diabatic Hamiltonian
\beqa
\hat{H}_{\rm CD}'=\hat{H}+\hat{H}_1'  =\hat{H}+ i\hbar\sum_n|\partial_tn\ra\la n|.
\eeqa
The experimental demonstration of the CD technique has recently been reported in effective two-level systems arising in a Bose-Einstein condensate trapped in an optical lattice \cite{expCD1} as well as in an electron spin of a single nitrogen-vacancy center \cite{expCD2}. It can be extended to many-body spin systems undergoing a quantum phase transition, at the cost of implementing $n$-body interactions \cite{DRZ12}, which are necessary to suppress the universal formation of excitations and defects \cite{QKZM}.  
Nonetheless, the application of this technique to systems with continuous variables such as matter waves seems to be hindered by the non-local nature of the counter-diabatic Hamiltonian $\hat{H}_1'$. Jarzynski has recently shown that for a quantum piston with one degree of freedom $q$ and for any power-law trap of the form $V(q)=\alpha |q|^{b}$ \cite{Jarzynski13}, the counter-diabatic term is
\beqa
\hat{H}_1' \propto (qp+pq),
\eeqa
where $p$ is the momentum operator canonically conjugated to $q$.
 This is in agreement with previous calculations for a time-dependent harmonic trap  ($b=2$) \cite{Muga10b}.
The implementation of this non-local Hamiltonian in trapped ions and ultracold atoms by optical means seems challenging.
Another important limitation of the CD technique is that it demands knowledge of the spectral properties of the Hamiltonian $\hat{H}(t)$, that is, of the instantaneous eigenvalues and eigenstates as a function of $t$. 
As a result, the CD is unsuited for systems obeying an effectively-nonlinear dynamics that often arises in mean-field/Hartree-Fock approximations, as it is the case for Bose-Einstein condensates
described by the Gross-Pitaevskii equation and its variants \cite{Ueda}. Further, direct computation of $\hat{H}_1$ ($\hat{H}_1'$) in other many-body systems is either impossible due to the lack of knowledge of the spectral properties, or would constitute a daunting task as in the case of exactly solvable models, e.g. by Bethe ansatz in homogeneous potentials.

In this letter, all these difficulties are overcome
by reformulating counter-diabatic driving in terms of scaling laws.
Specifically,  i) a  counter-diabatic driving protocol is derived  for single-particle, many-body, and non-linear  systems undergoing self-similar dynamics in a variety of trapping potentials,
ii)  the requirement to diagoanalize the Hamiltonian of the system $\hat{H}(t)$ is completely removed, 
iii ) it is shown that while the requirement of a non-local counter-diabatic term of the form $\hat{H}_1'\propto (qp+pq)$ is a common feature in matter waves (such as trapped ions, ultracold gases and other strongly-correlated quantum fluids), it is possible to find an alternative representation of $\hat{H}_{\rm CD}$ associated with an auxiliary term $\hat{H}_1$ which is local and experimentally realizable with well-established techniques.



Let us consider the  broad family of many-body systems, described by the Hamiltonian
\beqa
\label{mbh}
\hat{\mathcal{H}}&=&\!
\sum_{i=1}^N\!\Big[-\frac{\hbar^2}{2m}\Delta_{\q_i}
+\frac{1}{2}m\om^2(t)\q_i^2+\U(\q_i,t)\!\Big]\!\nonumber\\
& & +\epsilon(t)\sum_{i<j}V(\q_i-\q_j),
\eeqa
where ${\q_i}\in\mathbb{R}^D$ ($D$ denoting the effective dimension of the system), $\Delta_{\q_i}$ is the Laplace operator, and $\U(\q_i,t)$ represents an external trap  whose time-dependence is of the form $\U(\q,t)=\U(\q/\gamma,0)/\gamma^2$, $\gamma=\gamma(t)$ being a function of time. 
The two-body interaction potential obeys 
\beqa
{\rm V}(\lambda \q)=\lambda^{-\alpha}{\rm V}(\q),
\eeqa
which among other  relevant examples includes the  pseudo-potential describing s-wave scattering in ultracold gases, for which $\alpha=D$.
Without loss of generality, we choose the dimensionless time-dependent coupling constant $\epsilon(t)$ to satisfy $\epsilon(0)=1$. 
We further consider a stationary state at $t=0$, $\Phi(t)=\Phi(\q_1,\dots,\q_N;t)$, with chemical potential $\mu$, i.e.,  $\hat{\mathcal{H}}\Phi=\mu\Phi$ 
(an assumption we shall remove below).

We begin by pointing out that for a single-particle harmonic oscillator of a single-degree of freedom,  
two-eigenstates $|n(\om_1)\ra$ and $|n(\om_2)\ra$ corresponding to the  $n$-th mode of two different traps of frequencies $\om_1$ and $\om_2$ 
are related by the scaling transformation
\beqa
\la q |n(\om_2)\ra=\gamma_{12}^{-1/2}\la q/\gamma_{12}|n(\om_1)\ra,
\eeqa
where $\gamma_{12}=[\om_1/\om_2]^{\frac{1}{2}}$ plays the role of a (adiabatic) scaling factor.  
Under such type of scaling, the dynamics is self-similar in real-space.
An analogous scaling law, is fulfilled in the quantum piston and power-law traps recently discussed by Jarzynski \cite{Jarzynski13}, and can be extended to  many other systems \cite{GBD10}.
Indeed, exploitation of scaling laws has proved useful in engineering a type of shortcut to adiabatic which does {\it not} drive the dynamics through the adiabatic manifold of 
$\hat{\mathcal{H}}(t)$ \cite{delcampo11b,DB12}.
Motivated by this observation, we consider the time-evolution $\Phi(t)$ of the  initial  state to be described by the simple scaling
\beqa
\label{mbphit}
\Phi(t)=\gamma^{-\frac{ND}{2}}
e^{-i\mu \tau(t)/\hbar}
\Phi\Big[\frac{\q_1}{\gamma(t)},\dots,\frac{\q_N}{\gamma(t)};0\Big],
\eeqa
where the prefactor accounts for the correct normalization and $\tau(t)$ is a function of time soon to be determined.
We note that we will not necessitate knowledge of the precise form of $\Phi(t=0)$.
Direct substitution into the many-body Schr\"odinger equation  for $\hat{\mathcal{H}}$ shows that 
the ansatz Eq. (\ref{mbphit}) is actually the exact time-dependent solution of the counter-diabatic Hamiltonian $\hat{\mathcal{H}}'_{\rm CD}$, with 
\beqa
\label{mbh}
\gamma^2 \hat{\mathcal{H}}'_{\rm CD}&=&\!
\sum_{i=1}^N\!\Big[-\frac{\hbar^2}{2m}\Delta_{\s_i}
+\frac{1}{2}m\om^2(t)\gamma^4\s_i^2+\U(\s_i,0)\nonumber\\
& & -i\frac{\hbar \partial_{\tau}\gamma}{2\gamma}(\s_i\partial_{\s_i}+\partial_{\s_i}\s_i)\Big]\!
+\epsilon\gamma^{2-\alpha}\sum_{i<j}V(\s_i-\s_j),\nonumber\\
\eeqa
where the scaled spatial coordinate and time variables read
\beqa
\label{scaledvars}
\s_i=\q_i/\gamma ,\qquad \tau=\int^t\gamma^{-2}(t')dt'.
\eeqa 
One can choose a suitable time-dependence of $\om^2(t)$ and $\epsilon(t)$ to render $\gamma^2 \hat{\mathcal{H}}'_{\rm CD}$ 
 time-independent, and to reduce it to its form at $t=0$, when $\om(0)=\om_0$ and $\epsilon(0)=1$.
This requirement leads to the the consistency conditions
\beqa
\label{gamma}
\gamma(t)&=&\bigg[\frac{\om_0}{\om(t)}\bigg]^{\frac{1}{2}},\\
\label{epsilon}
\epsilon(t)&=&\bigg[\frac{\om_0}{\om(t)}\bigg]^{\frac{\alpha-2}{2}}.
\eeqa
Under this choice, $\Phi(0)$ remains a stationary solution of $\gamma^2\hat{\mathcal{H}}'_{\rm CD}$ with chemical potential $\mu$.
Similarly, an initial nonstationary state follows a trivial dynamics in the scaled variables \cite{note}.
The first consistency equation can be considered  the definition of the adiabatic scaling factor in a self-similar dynamics.
It agrees with the well-known result of the single-particle harmonic oscillator.
The second condition implies that for a self-similar dynamics to occur in an interacting-many-body system 
it might be necessary to change along the process the amplitude of the interactions.
In ultracold gases, this can be achieved by means of a Feschbach resonance or a modulation of the transverse confinement. 
However, for moderate changes of the scaling factors, this condition can be dropped  -one can set $
\epsilon(t)=1$- and the dynamics is self-similar  to a high accuracy \cite{DB12}.
In addition,  certain systems satisfy the condition $\alpha=2$ so that $\epsilon(t)=1$ is fulfilled exactly, and no interaction tuning is required.
This is the case of a two-dimensional Bose gas, as a reflection of the Pitaevskii-Rosch symmetry \cite{PR97}. The same is true
 for the family of Tonks-Girardeau gases with effectively infinite strength of interactions \cite{GNO04}, the Calogero-Sutherland model \cite{Sutherland98}, etc.

Provided that these equations are satisfied, the scaling ansatz (\ref{mbphit}) is an exact solution of the many-body time-dependent Schr\"odinger equation.
This result provides a way to apply counter-diabatic driving to the broad family of systems described by Eq. (\ref{mbh}) without the need to perform the explicit computation in (\ref{hCD}).
By doing so, it removes the requirement to diagonalize the Hamiltonian of the system of interest $\hat{H}(t)$, 
a feature particularly useful to design shortcuts to adiabaticity in interacting many-body systems.
As we shall discuss below, it can be applied as well to non-linear equations of motion.
Further, one can identify the auxiliary CD term  
\beqa
\label{H1p}
\hat{\mathcal{H}}_1'=-i\frac{\hbar \dot{\gamma}}{2\gamma}\sum_{i=1}^N(\q_i\partial_{\q_i}+\partial_{\q_i}\q_i), 
\eeqa
which is needed to drive an {\it arbitrary} self-similar dynamics of matter-waves through the adiabatic manifold $\hat{\mathcal{H}}$ 
(dots denote derivatives with respect to time $t$ throughout the text).

A non-local term like $\hat{\mathcal{H}}_1'$  is present in generalized harmonic oscillators, and following \cite{bookinv},  
we next find an alternative representation of $\hat{\mathcal{H}}_{\rm CD}'$ in which this term is absent. 
In order to do so, we consider the $N$-body canonical transformation,
\beqa
\mathcal{U}
&=&\prod_{i=1}^N\exp\left(\frac{im \dot{\gamma}}{2\hbar \gamma}\q_i^2\right)
\eeqa
whose action on  $\q_i$, $\p_i$, and $\hat{\mathcal{H}}_{\rm CD}'$ is as follows,
\beqa
\q_i&\rightarrow& \mathcal{U}\q_i\mathcal{U}^{\dag}=\q_i,\\
\p_i&\rightarrow& \mathcal{U} \p_i\mathcal{U}^{\dag}=\p_i -\frac{m \dot{\gamma}(t)}{\gamma(t)}\q_i, \label{pshift}\\
\hat{\mathcal{H}}_{\rm CD}' &\rightarrow& \hat{\mathcal{H}}_{\rm CD}(t)=\mathcal{U}\hat{\mathcal{H}}_{\rm CD}'(t)\mathcal{U}^{\dag}-i\hbar \mathcal{U}\partial_t\mathcal{U}^{\dag}.
\eeqa
As a result of (\ref{pshift}),  the kinetic energy term in the Hamiltonian is modified, and upon expansion the non-local counter-diabatic term (\ref{H1p}) is cancelled.
The new representation $\hat{\mathcal{H}}_{\rm CD}(t)$ of the counter-diabatic Hamiltonian  resembles that of the original Hamiltonian   $\hat{\mathcal{H}}$,
\beqa
\label{mbCDhmapped}
\hat{\mathcal{H}}_{\rm CD}&=&\!
\sum_{i=1}^N\!\Big[-\frac{\hbar^2}{2m}\Delta_{\q_i}
+\frac{1}{2}m\Om^2(t)\q_i^2+U(\q_i,t)\Big]\!\nonumber\\
& & +\epsilon(t)\sum_{i<j}V(\q_i-\q_j),
\eeqa
 with the modified time-dependent (squared) frequency
\beqa
\label{Omeff}
\Om^2(t)&=&\om^2(t)- \frac{\ddot{\gamma}}{\gamma},
\eeqa
this is, the counter-diabatic driving reduces the trap frequency by a term proportional to the acceleration of the scaling factor, 
a result which resembles the outcome of the Duru transformation in transport problems \cite{Duru89}.
For $\gamma(t)$ in (\ref{gamma}), Eq. (\ref{Omeff})  reduces to 
\beqa
\Omega^2(t)=\omega^2(t)-\frac{3}{4}\bigg[\frac{\dot{\omega}(t)}{\omega(t)}\bigg]^2
+\frac{1}{2}\frac{\ddot{\omega}(t)}{\omega(t)}.
\label{omeff}
\eeqa
Under this canonical transformation the time-evolution of the initial state, instead of being described by Eq. (\ref{mbphit}), is mapped to $\Phi(t) \rightarrow \Psi(t)=\mathcal{U}\Phi(t) $, with
\beqa
\label{psit}
\Psi(\q_1,\dots,\q_N;t)=
\exp\left(
-i\frac{m\dot{\om}(t)}{4\hbar\om(t)}\sum_{i=1}^N\q_i^2\right)\Phi(t),
\eeqa
 that is, it acquires the Berry phase which vanishes only as $\dot{\omega}/\omega\rightarrow 0$ but is non-zero along a shortcut to adiabaticity assisted by CD.

Let us summarize the results above.
To drive the dynamics of a given state of $\hat{\mathcal{H}}$ mimicking adiabaticity for the driving $\om(t)$, 
it suffices to implement instead the counter-diabatic driving $\Om(t)$ and to fulfill the consistency conditions Eqs. (\ref{gamma}) and (\ref{epsilon}), 
which  are still referred to the driving $\om(t)$.
It is worth pointing out that the frequency modulation in Eq. (\ref{omeff}) was derived before in a different context, 
that of a single-particle time-dependent harmonic oscillator, as the consistency condition for a time-dependent Gaussian variational ansatz for the instantaneous ground state \cite{delcampo11}. 
CD for self-similar process is as well related to the inversion of the scaling laws for a time-dependent harmonic oscillator \cite{Chen10}, 
choosing $\gamma(t)$ according to Eq. (\ref{gamma}) and promoting $\om(t)\rightarrow\Om(t)$ in the corresponding consistency equation (Ermakov equation).

A possible implementation of a shortcut to adiabaticity assisted by CD would be as follows.
Consider the case in which we are interested in driving 
a transition from $|n(0)\ra$ to $|n(t_{\rm F})\ra$ in a time $t_{\rm F}$. 
The condition for the CD driving Hamiltonian to equal the system Hamiltonian $\hat{\mathcal{H}}_{\rm CD}(t)=\hat{\mathcal{H}}(t)$ at $t=\{0,t_{\rm F}\}$, leads to the boundary conditions
\beqa
\label{BC}
\om(0)&=&\om_0, \dot{\om}(0)=0,  \ddot{\om}(0)=0,\nonumber\\
\om(t_{\rm F})&=&\om_F, \dot{\om}(t_{\rm F})=0,  \ddot{\om}(t_{\rm F})=0,
\eeqa
which can be satisfied by  an interpolating ansatz. Choosing it to be a polynomial $\om(t)=\sum_{k=1}^6a_k(t/t_{\rm F})^{k-1}$ 
we find 
\beqa
\label{omt}
\om(t)=\om_0+10\delta s^3-15\delta s^4+6\delta s^5,
\eeqa
where $\delta=\om_{\rm F}-\om_0$ and $s=t/t_{\rm F}$  for short.

\begin{figure}
\centering{\includegraphics[width=0.95\linewidth]{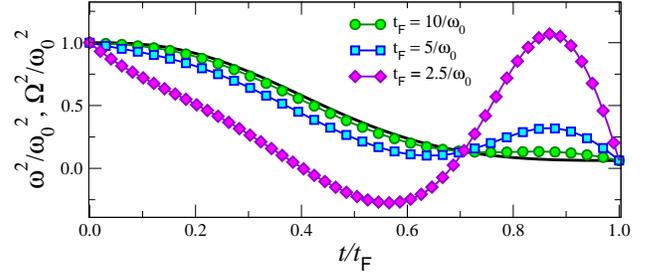}}
\caption{
{\bf  Shortcut to adiabaticity by counter-diabatic driving.}
The time-dependent  frequency $\om(t)$ (solid line) in Eq. (\ref{omt}) does not suffice to fulfill adiabaticity in a finite time $t_{\rm F}$.
The necessary counter-diabatic driving of the trap frequency $\Om(t)$ for a shortcut to an adiabatic expansion is given by
Eq. (\ref{omeff})  and is displayed here for three different values of $t_{\rm F}$ (symbols), 
for an expansion factor $\gamma(t_{\rm F})=2$. The values of the product $\om_0t_{F}$ correspond 
to the onset of the adiabatic limit ($\circ$), a moderate speed-up ($\Box$),  and an ultrafast expansion ($\diamond$),   
}
\label{omtfig}
\end{figure}

Using Eq. (\ref{omeff}), the required modulation of the trap frequency $\Om^2(t)$ to implement a shortcut to adiabaticity in a finite time $t_{\rm F}$ can be derived.
Instances of the resulting counter-diabatic driving are displayed in Fig. \ref{omtfig}.
For sufficiently slow expansions, $\Om^2(t)$ approaches $\om^2(t)$,  as expected.
For moderate speed-ups, the non-monotonic  behavior of  $\Om^2(t)$ as  function of $t$ is enhanced. 
In the limit of exceedingly short expansion times, $\Om^2(t)$ can become temporarily negative. 
This is tantamount to the transient inversion of the trapping potential during which it becomes an expelling barrier, providing the required speed up. 
Such modulation can be implemented by applying an offset field in optical traps \cite{Khaykovich02} or using time-averaging potentials \cite{painters}. 
For expansion times  such that $\Om^2(t)>0$, the counter-diabatic driving can be implemented as in the experiments with ultracold gases and Bose-Einstein 
condensates reported by Schaff {\it et al.} where a magnetic confinement was used \cite{Schaff1,Schaff2}. 

In the following, we shall illustrate the usefulness of this new formulation of the counter-diabatic driving 
by engineering a shortcut to adiabaticity in systems where the design based on Eq. (\ref{hCD}) is not applicable.
This is the case for the effective non-linear dynamics of Bose-Einstein condensates, which precludes the application of the superposition principle, 
exploited to derive the form of $\hat{\mathcal{H}}_{\rm CD}$ in Eq. (\ref{hCD}).
Within a mean-field description, the time-dependent Gross-Pitaevskii equation (TDGPE)  
\beqa
i\hbar\partial_t\Phi(\q,t)&=&\big[-\frac{\hbar^2}{2m}\Delta_{\q}+\frac{1}{2}m\om^2(t)\q^2\nonumber\\
& & +\U(\q,t)+g_{D}|\Phi(\q,t)|^2\big]\Phi(\q,t),
\eeqa
rules the dynamics of  $\Phi(\q,t)$, the normalized condensate wavefunction.
For a self-similar dynamics of the form $\Phi(\q,t)=\gamma^{-\frac{D}{2}}\exp(-i\mu\tau(t)/\hbar)\Phi(\q/\gamma,t=0)$ the non-local auxiliary term 
$\hat{\mathcal{H}}_1'=-i\frac{\hbar \dot{\gamma}}{2\gamma}(\q\partial_{\q}+\partial_{\q}\q)$ is required.
Alternatively, $\Psi(t)=\exp\left(i\frac{m|\q|^2\dot{\gamma}}{2\gamma\hbar}\right)
\Phi(\q,t)$
satisfies the TDGPE provided that $\tau(t)$, $\gamma(t)$ and $\Om^2(t)$ are given by 
Eqs. (\ref{scaledvars}), (\ref{gamma}), and (\ref{omeff}) respectively, and that $g_D(t)=g_D(0)\gamma^{D-2}$.
$\Psi(t)$ differs from the previous scaling laws reported for Bose-Einstein condensates \cite{CDKSS} in that the scaling factor 
$\gamma(t)$ follows the adiabatic trajectory.
Note however that the same protocol holds exactly beyond mean-field, that is, it induces the time evolution of the many-body wavefunction according to Eq. (\ref{psit}) 
for the microscopic model in Eq. (\ref{mbCDhmapped}) with the Fermi-Huang pseudopotential ($V(\q)=g\delta(\q)d_{|\q|}(|\q|\cdot)$).
Many experiments are performed in the the Thomas-Fermi (TF) regime, where the mean-field energy dominates over the kinetic energy contribution, and the term with the Laplacian can be dropped. Remarkably, it is then possible to find an exact counter-diabatic driving protocol without tuning the interaction strength, so that $g(t)=g(0)$, in any dimension $D$. 
Redefining $\tau(t)=\int^t\gamma(t')^{-D}dt'$, it is found that the counter-diabatic frequency takes the form 
\beqa
\Om_{\rm TF}^2(t)=\frac{\om_0^2}{\gamma^{D+2}}-\frac{\ddot{\gamma}}{\gamma},
\eeqa 
which results in
\beqa
\Om_{\rm TF}^2(t)=\om_0\om(t)\bigg[\frac{\om(t)}{\om_0}\bigg]^{\frac{D}{2}}-\frac{3}{4}\bigg[\frac{\dot{\omega}(t)}{\omega(t)}\bigg]^2+\frac{1}{2}\frac{\ddot{\om}(t)}{\om(t)}.
\label{omeffTF}
\eeqa
If no auxiliary term is to be present at the beginning and end of the process, 
 $\om(t)$ has to satisfy the boundary conditions in (\ref{BC}), as it is the case for the polynomial ansatz in Eq. (\ref{omt}). Implementation of the driving frequency $\Om_{\rm TF}(t)$ in the laboratory (e.g.  as in \cite{Schaff2}), would induce a fast motion video of the adiabatic dynamics of the condensed gas. Using the recently developed non-destructive Faraday imaging, 
it would be even possible to track the evolution of the cloud, and to measure $\gamma(t)$ in a single experimental realization \cite{Sherson13}. 
This would constitute a remarkable demonstration of the shortcuts proposed here.

It should be clear that our results are directly applicable to other self-similar evolutions, in the absence of a harmonic trap in $\hat{\mathcal{H}}$ (i.e., with $\om(t)=0$), as it is the case for the dynamics in time-dependent power-law potentials recently discussed by Jarzynski \cite{Jarzynski13}. We consider here the extension to an arbitrary dimension $D$ and  power exponent $b$, ${\rm U}(\q,t)=\alpha|\q/\xi(t)|^b$, where $\alpha>0$ and $\xi(t)>0$ is the time-dependent width.
Noticing that the scaling factor is then given by $\gamma(t)=[\xi(t)/\xi(0)]^{\frac{b}{b+2}}$, it follows that the local counter-diabatic driving  associated with  the single-particle shortcuts proposed in \cite{Jarzynski13}, corresponds to the auxiliary potential 
\beqa
\hat{\mathcal{H}}_1=-\frac{1}{2}\frac{b}{b+2}m\frac{\ddot{\xi}}{\xi}\q^2,
\eeqa
which is applicable to a variety of many-body systems as well (those for which $\alpha=2$), and in particular to Bose-Einstein condensates in the Thomas-Fermi regime. For systems in Eq. (\ref{mbh}) with $\alpha\neq 2$, the self-similar dynamics is to be assisted by tuning the strength of the interactions according to $\epsilon(t)=\gamma(t)^{\alpha-2}$. 
The case $b\rightarrow\infty$ corresponds to an expanding quantum piston \cite{DB12,Stefanatos13,Jarzynski13}, where the external trap is a time-dependent box of width $\xi(t)$, and the counterdiabatic-term reduces precisely to  the auxiliary potential $\U^{\rm aux}(\q)=-\frac{1}{2}m\frac{\ddot{\xi}}{\xi}\q^2$ engineered in \cite{DB12}. 

Before closing, we discuss an important application at the single-particle level of relevance to thermal ultracold gases, and in particular, to trapped ions.
A realization of a quantum engine has been proposed using an Otto cycle with a single ion in a time-dependent trap as a working medium \cite{Huber08,Abah12}.
Using shortcuts to adiabaticity, it is possible to engineer a ``superadiabatic engine'' which operates at maximum efficiency and has tunable output power \cite{DGP13}.
The driving protocol proposed here allows one to achieve this goal by substituting the standard compression and expansion adiabats by their speeded up counterparts. 

%
In conclusion, a reformulation of the counter-diabatic driving technique has been presented which allows the  engineering of shortcuts to adiabaticity
 in a large class of single- and many-particle quantum systems confined in time-dependent traps. In doing so, we have removed the requirement 
to diagonalize the instantaneous Hamiltonian as well as the restriction to linear dynamics. 
The auxiliary counter-diabatic term needed to speed up an arbitrary self-smilar dynamics has been found in all these systems. 
The resulting  protocol  drives the evolution along the adiabatic 
trajectory of the system of interest and it involves a time-dependent harmonic potential which can be implemented with well-established  
experimental techniques. Extensions of the formulation of counter-diabatic driving presented in this work can be envisioned to account for other self-similar processes associated with the conformal symmetry \cite{Galajinsky08}.


\begin{acknowledgments}
The author would like to thank  
E. Passemar, 
D. Roy, and N. Sinitsyn for insightful discussions. 
This work is supported by the U.S. Department of Energy through the LANL/LDRD Program and a  LANL J. Robert Oppenheimer fellowship.
\end{acknowledgments}


\begin{thebibliography}{99}
\bibitem{note} The result clearly applies to  an initial state described by a linear superposition $|\Phi(0)\ra=\sum_\nu c_\nu|\Phi_\nu(0)\ra$
of many-body eigenstates $|\Phi_\nu\ra$ of energy $\varepsilon_\nu$, where $\nu$ denotes a complete set of quantum numbers.  
The evolution generated by $\mathcal{H}_{\rm CD}'$ leads to $|\Phi(t)\ra=\sum_\nu c_\nu\exp[-i\varepsilon_\nu\tau(t)/\hbar]|\Phi_\nu(t)\ra$. 
It is equally applicable to  mixed states  $\rho=\sum_\nu p_\nu|\Phi_\nu\ra\la \Phi_\nu|$.
\end{thebibliography}
\end{document}